\documentclass[10pt,journal,compsoc]{IEEEtran}

\ifCLASSOPTIONcompsoc
  \usepackage[nocompress]{cite}
\else
  \usepackage{cite}
\fi

\ifCLASSINFOpdf
\else
\fi

\usepackage{graphicx}
\usepackage{rotating}
\usepackage{booktabs} 
\usepackage{hyphenat}
\usepackage{paralist}

\usepackage{amsmath} 
\usepackage{url} 

\usepackage{balance}
\usepackage{microtype}

\usepackage{colortbl}

\usepackage{tabularx}

\usepackage{listings}

\usepackage{subcaption}

\definecolor{lightgray}{gray}{0.75}

\usepackage{float}

\usepackage[T1]{fontenc} 
\usepackage{pifont}
\usepackage{calc}
\usepackage{xcolor}
\usepackage{pdfpages}

\usepackage{wasysym}
\usepackage{tikz}
\usepackage[scaled]{beramono}

\usepackage{lipsum}

\usepackage{framed}

\usepackage{amssymb}

\usepackage{romannum}

\usepackage{listings}

\lstdefinestyle{java}{
	language=Java,
	tabsize=4,
	breaklines=false,
	basicstyle=\fontfamily{pcr}\scriptsize\selectfont,
	commentstyle=\fontshape{it}\color{comment2}\selectfont,
	keywordstyle=\fontseries{b}\selectfont,
	stringstyle=\fontfamily{pcr}\selectfont,
	numbers=none,
	numberstyle=\footnotesize,
	captionpos=b,
	frame=tb,
	framesep=3pt,
	xleftmargin=4pt,
	xrightmargin=4pt,
	rulecolor=\color{bgborder},
	firstnumber=auto,
}

\lstdefinestyle{xml}{
	style=java,
	tabsize=2,
	language=xml,
}

\lstdefinestyle{featureidexml}{
	style=xml,
	morekeywords={featureModel, struct, feature, name, abstract, mandatory, and, alt, or, constraints, rule, imp, var, comments},
}

\lstdefinestyle{jak}{
	style=java,
	morekeywords={layer, refines, Super(), Super},
}

\lstdefinestyle{featurehouse}{
	style=java,
	morekeywords={original, refines},
}

\lstdefinestyle{deltaj}{
	style=java,
	morekeywords={original, core, delta, when, after, before, modifies, adds, removes},
}

\lstdefinestyle{aspectj}{
	style=java,
	morekeywords={aspect, after, call},
}

\lstdefinestyle{ifdef}{
  morecomment=[l]{\#},
}

\lstdefinestyle{interface}{
	morekeywords={requires, provides},
}

\lstdefinestyle{jml}{
	commentstyle=\fontshape{it}\color{green3}\selectfont,
}

\lstdefinestyle{tiny}{
	basicstyle=\fontfamily{pcr}\tiny\selectfont
}

\lstdefinestyle{scriptsize}{
	basicstyle=\fontfamily{pcr}\scriptsize\selectfont
}

\lstdefinelanguage{Coq}{
  morekeywords={Variable,Inductive,CoInductive,Fixpoint,CoFixpoint,
      Definition,Lemma,Theorem,Axiom,Local,Save,Grammar,Syntax,Intro,
      Trivial,Qed,Intros,Symmetry,Simpl,Rewrite,Apply,Elim,Assumption,
      Left,Cut,Auto,Unfold,Exact,Right,Hypothesis,Pattern,Destruct,
      Constructor,Defined,Fix,Record,Proof,Induction,Hints,Exists,let,in,
      Parameter,Split,Red,Reflexivity,Transitivity,if,then,else,Opaque,
      Transparent,Inversion,Absurd,Generalize,Mutual,Cases,of,end,Analyze,
      AutoRewrite,Functional,Scheme,params,Refine,using,Discriminate,Try,
      Require,Load,Import,Scope,Set,Open,Section,End,match,with,Ltac,
			exists,forall
	},
  sensitive, 
  morecomment=[n]{(*}{*)},
  morestring=[d]",
  literate=
   {=>}{{$\Rightarrow$}}1
	 {>->}{{$\rightarrowtail$}}2
	 {<->}{{$\leftrightarrow$}}2
	 {->}{{$\to$}}1
    {\/\\}{{$\wedge$}}1
    {|-}{{$\vdash$}}1
    {\\\/}{{$\vee$}}1
    {~}{{$\sim$}}1
}[keywords,comments,strings]

\lstdefinestyle{coq}{
	language=Coq,
	tabsize=4,
	breaklines=false,
	columns=[c]fixed,
	flexiblecolumns=false,
	basicstyle=\fontfamily{pcr}\footnotesize\selectfont,
	commentstyle=\color{coqcomment}\fontshape{it}\selectfont,
	keywordstyle=\color{coqblue}\fontseries{b}\selectfont,
	stringstyle=\fontfamily{cmr}\selectfont,
	numbers=none,
	numberstyle=\footnotesize,
	captionpos=b,
	frame=single,
	xleftmargin=4pt,
	xrightmargin=4pt,
	rulecolor=\color{bgborder},
	morekeywords={Module, Fact, Export, Tactic, Notation, Prop, Combined, Variables},
  emph={Variable, Hypothesis, Module, Fact, Lemma, Theorem, Ltac, Inductive, Fixpoint, Definition, Parameter, with, admit, Variables},
	emphstyle={\color{coqred}\fontseries{b}\selectfont},
  escapeinside={$}{$}
}

\lstdefinestyle{slides}{
	backgroundcolor=\color{white}
}

\restylefloat{table}
\renewcommand{\paragraph}[1]{\vspace{.4em}\noindent\textbf{#1.} } 

\definecolor{gray}{rgb}{0.87,0.87,0.87}

\newlength{\dinglength}
\makeatletter
\newcommand\HL{
	\gdef\lst@alloverstyle##1{
		\fboxrule=0pt
		\fboxsep=0pt
		\colorbox{gray}{\strut##1}
	}
}
\newcommand\HLoff{
	\xdef\lst@alloverstyle##1{##1}
}

\newcommand\samplingevidenceforlowdegree{ABW:ASE14,CCR:SPLC10,CDS:ISSTA07,GC:ISSRE11,KWG:TSE04,NL:CSUR11,MKRGA:ICSE16}

\begin{document}
	
\graphicspath{{img/}}
	\title{Understanding Differences among Executions with Variational Traces}

\author{Jens Meinicke,{\large$^{1,2}$} Chu-Pan Wong,{\large$^1$} Christian K{\"a}stner,{\large$^1$} Gunter Saake{\large$^2$}
	\\[2pt]
	{\small $^1$Carnegie Mellon University, USA, $^2$University of Magdeburg, Germany}
}

{

\IEEEtitleabstractindextext{
\begin{abstract}
One of the main challenges of debugging is to understand why the program fails for certain inputs but succeeds for others.
This becomes especially difficult if the fault is caused by an interaction of multiple inputs.
To debug such interaction faults, it is necessary to understand the individual effect of the input, how these inputs interact and how these interactions cause the fault.
The differences between two execution traces can explain why one input behaves differently than the other.
We propose to compare execution traces of all input options to derive explanations of the behavior of all options and interactions among them.
To make the relevant information stand out, we represent them as variational traces that concisely represents control-flow and data-flow differences among multiple concrete traces.
While variational traces can be obtained from brute-force execution of all relevant inputs, we use variational execution to scale the generation of variational traces to the exponential space of possible inputs. 
We further provide an Eclipse plugin Varviz that enables users to use variational traces for debugging and navigation.
In a user study, we show that users of variational traces are more than twice as fast to finish debugging tasks than users of the standard Eclipse debugger.
We further show that variational traces can be scaled to programs with many options.\looseness=-1
\end{abstract}

\begin{IEEEkeywords}
Debugging, Program Comprehension, Feature Interaction, Configurable Software, Variational Execution.
\end{IEEEkeywords}}

\maketitle

\IEEEraisesectionheading{\section{Introduction}}\noindent
\noindent
Understanding why a certain program input causes a fault while another succeeds is a common task during debugging~\cite{Z:FSE99}.
This happens, for example, if a certain program crashes in one configuration but succeeds if a parameter is changed~\cite{CDS:ISSTA07, MKRGA:ICSE16, NL:CSUR11}.
Reasoning about such differences in executions is difficult when using a standard debugger as the program can only be executed for one input at a time.
To understand why certain inputs lead to a fault requires to understand the differences between the valid and the invalid executions.
Such a comparison can be generated by recording and aligning the traces and state changes of the two executions~\cite{XSZ:PLDI08, SZ:FSE10}.
The aligned traces can be used to generate explanations of the fault~\cite{SZ:ICSE13, Z:FSE02}.\looseness=-1

Some faults, however, are caused by interactions of multiple inputs which make understanding and debugging them even more challenging (aka. \emph{feature interaction})~\cite{ABW:ASE14, AKS+:FOSD13, GC:ISSRE11, CKMR03}.
Such interaction faults are common in practice, especially in highly configurable systems with many options~\cite{ABW:ASE14}, or if a set of changes introduces the fault~\cite{ZH:TSE02, BOR:FSE13, GBE+:ICSE10}.
Interaction faults are hard to detect as they require to specify a certain input to trigger the fault~\cite{MKR+2015}.
Even if we can narrow down the fault to a smaller number of options, say with \emph{delta debugging}~\cite{ZH:TSE02, Z:FSE99}, it is still difficult to understand \emph{why}, \emph{where}, and \emph{how} they interact.

\begin{figure*}[t]
	\begin{subfigure}[b]{0.24\textwidth}
		\begin{lstlisting}[tabsize=2, basicstyle=\fontsize{6}{7}\selectfont]
int netpollSetup(boolean ipv4, 
                 boolean flag) {
	Integer err = null;
	if (flag) {
		err = -1;
	}
	return getErrValue(err, ipv4);
}
int getErrValue(Integer err,
                boolean ipv4) {
	if (ipv4) {
		return err.intValue();
	}
	return 1;
}
		\end{lstlisting}
		Debug: \texttt{netpoll(?, ?);}
		
		\vspace{118pt}
		\caption{Source code}
		\label{fig:code}
	\end{subfigure}
	\begin{subfigure}[b]{0.17\textwidth}
		\includegraphics[trim=10pt 45pt 740pt 0pt,clip,page=1, width=1\textwidth]{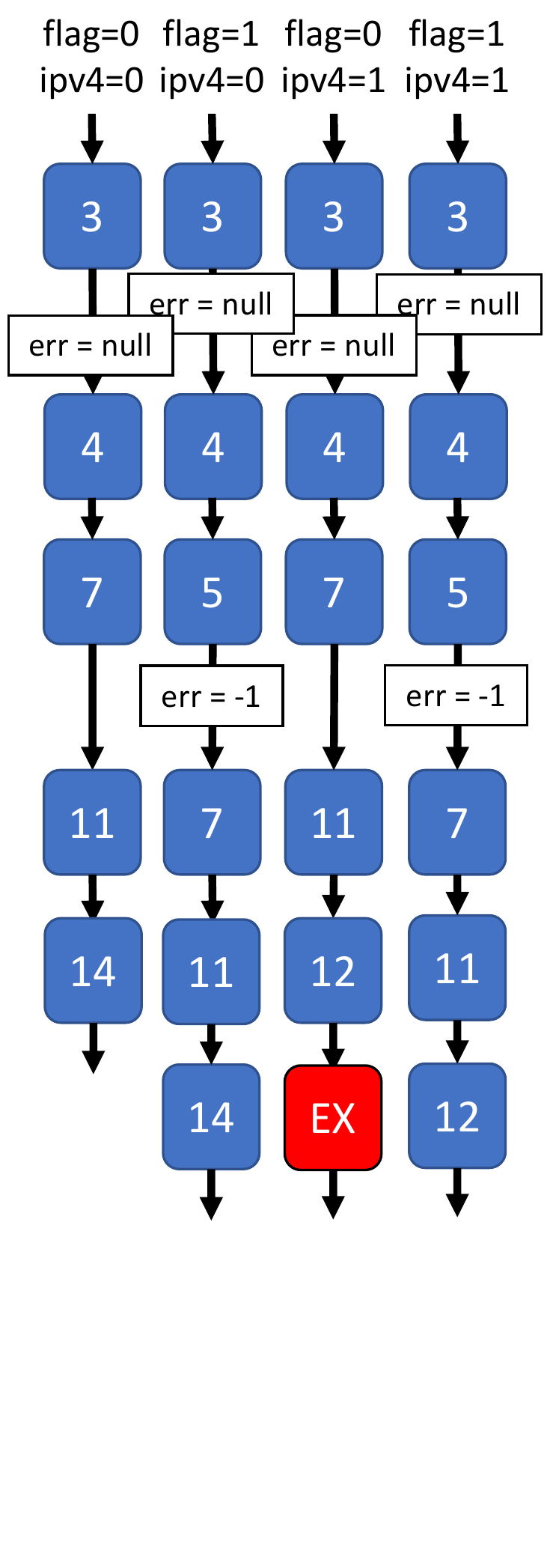}
		\caption{Recording}
		\label{fig:record}
	\end{subfigure}
	\begin{subfigure}[b]{0.17\textwidth}
		\includegraphics[trim=10pt 45pt 740pt 0pt,clip,page=2, width=1\textwidth]{traceNetpoll3.pdf}
		\caption{Alignment}
		\label{fig:align}
	\end{subfigure}
	\begin{subfigure}[b]{0.17\textwidth}
		\includegraphics[trim=10pt 45pt 740pt 0pt,clip,page=3, width=1\textwidth]{traceNetpoll3.pdf}
		\caption{Merging}
		\label{fig:merge}
	\end{subfigure}
	\begin{subfigure}[b]{0.223\textwidth}
		\includegraphics[trim=0pt 210pt 680pt 0pt,clip,page=4, width=1\textwidth]{traceNetpoll3.pdf}
		{\small State changes (orange rectangle), decisions (diamonds), their parameters (gray rectangle), the exception, and transitions that indicate the context of the execution.}

		\caption{Variational Trace}
		\label{fig:trace}
	\end{subfigure}
	\caption{Concept of a variational trace explaining the interaction fault caused by options flag and ipv4. Comparing executions requires to \emph{record} the executions for all inputs, \emph{align} the recorded traces, and \emph{merge} the aligned traces to identify interactions (the number on the elements indicate the line number). By enriching the merged trace with information about the state and control-flow decisions, a \emph{variational trace} can be created.}
	\label{fig:formal}
\end{figure*}

After identifying the set of interacting options, a programmer can start investigating how this interaction causes the fault.
Understanding the interaction requires understanding the individual behavior of the interacting options, but also their combined behavior.
Thus, it may no longer be sufficient to align the execution of two inputs as previous approaches do~\cite{Z:FSE02, SZ:ICSE13}, as such an alignment cannot explain the effects of multiple options.

We propose to align the execution traces of all configurations to explain the effect of multiple options.
We introduce \emph{variational trace}, a compact representation of the trace differences among all executions.
In the variational trace, redundant parts are shared and individual parts are annotated with the input they belong to.
This focus on differences allows understanding how data and control flow influence the executions and interact,
and thus how the different inputs cause a fault.
In Figure~\ref{fig:trace}, we show a variational trace that illustrates how the interaction fault of the example program is caused, as we will explain.

Generation of variational traces challenges scalability as the number of executions and traces that need to be aligned can potentially be exponential to the number of options.
A baseline approach would execute all configurations separately and thus can only scale to explain the interactions among few options.
More severely is the potential memory consumption as this approach needs to keep all past statements in memory, as it is upfront unclear which statements will differ among executions~\cite{KM:TOSEM10, PT:IEEES09}.

We use variational execution (a.k.a. faceted execution)~\cite{NKN:ICSE14, MWK+:ASE16, M14, AYF+:PLAS13} to avoid separately recording and aligning the traces for many inputs.
\emph{Variational execution} runs \emph{all} program configurations simultaneously, often efficiently, by sharing redundancies of the executions and values among these configurations.
Since the program is executed only once in a shared fashion, alignment is achieved by construction.
Additionally, executed statements and program states are tracked as they relate to the original inputs and their interactions.
This means that executions and variables can always be linked to specific program inputs.
With variational execution we also avoid the memory explosion as it enables us to decide on the fly which statements differ and need to be kept for the variational trace.\looseness=-1

To enable developers to interact with variational traces, we developed an interactive Eclipse plug-in called \emph{Varviz} that visualizes variational traces.
Varviz can be used for understanding faults, but also for other program comprehension tasks that involve understanding differences among similar executions.\looseness=-1

In a user study, we show that variational traces help to understand a fault in a highly interacting program and that they help to explain a fault from a previous experiment on automatic debugging techniques. 
Participants using variational traces were more than twice as fast and more successful compared to participants using a standard debugger.
In our qualitative study, we found that when dealing with faults caused by differences in inputs, in contrast to standard faults, participants search for places where these inputs interact and how they have an effect, which is exactly what our approach helps with.
Furthermore, we show that variational traces are fairly compact, even for medium sized programs with many explored options.\looseness=-1

\emph{The goal of this research is to aid researchers and developers detecting differences among executions depending on potentially multiple options by providing a dynamic analysis that is able to efficiently execute and align all configurations summarized in variational traces.}

Overall, the contributions of this paper are:
\begin{compactitem}
	\item The concept of a \emph{variational trace} that compactly represents differences among the executions of many configurations.\looseness=-1
	
	\item A \emph{baseline implementation} that shows challenges of generating variational traces.
		
	\item \emph{Efficient trace alignment} and a \emph{dynamic analysis to trace only relevant data} avoiding memory explosion for a potentially exponential number of configurations using  variational execution~\cite{M14, MWK+:ASE16}.
	
	\item An Eclipse plug-in \emph{Varviz} to visualize and interact with variational traces.
	
	\item A \emph{user study} that shows that participants using variational traces outperform participants using a standard debugger.
	 The study also shows that comparative approaches~\cite{SZ:ICSE13, Z:FSE02} actually help with debugging tasks.
	
	\item An \emph{evaluation on scalability} showing that our approach is able to align executions for large number of input combinations, while concisely describing their differences. 
	By focusing on the effects of certain inputs, the sizes can be reduced even further.
\end{compactitem}

\section{State of the Art}
\label{sec:motivation}
\noindent
An important problem of many program comprehension and debugging tasks is to understand the differences among executions.
For example, a programmer may want to understand why a fault occurs for certain inputs, but does not for others.
Such faults can be hard to detect, as they are only triggered for a certain input, and hard to understand because they may require reasoning about interactions among these inputs within
long traces.
The differences in control flow and data flow among faulty and valid executions can provide insights about how the fault is caused.

\begin{sloppypar} 
Our approach combines ideas from two research fields, namely \emph{automatic debugging} and \emph{feature interactions}, to explain differences among executions.
In this section, we discuss how automatic-debugging techniques exploit differences in executions and 
how various approaches address the feature-interaction problem that causes undesired behavior for certain combinations of inputs.\looseness=-1
\end{sloppypar}

\subsection{Automated Debugging Techniques}
\noindent
Automated-debugging techniques aim to create explanations of why a fault appears, by comparing valid and faulty executions of the program~\cite{JHS:ICSE02, Z:FSE02,WZSJ:ISSTA10, JCC+:SP11,SZ:ICSE13, SZ:FSE09}.
To create explanations, the approaches execute and record the program multiple times for different inputs or test cases.

\paragraph{Fault Localization}
There are many approaches that aim to find the cause of a fault~\cite{WGL+:TSE16}.
One of these approaches is \emph{spectrum-based fault localization} which rates each line of the source code by whether it is probable to cause the fault~\cite{JHS:ICSE02, AZG:PRDC06}.
For example, Tarantula compares the code coverage of valid and failing test cases and provides this information using different background colors on the code~\cite{JHS:ICSE02}.
However, such statement ranking has been shown to be less useful than expected~\cite{PO:ISSTA11}:
there are usually too many lines highlighted in the code, which makes it hard for users to understand which lines matter for an explanation of the fault.
Also, the information why certain lines are important is missing, as are control- and data-flow information.
Such information is however often necessary to understand how certain parts of the execution lead to the fault.\looseness=-1

\paragraph{Execution Comparison}
The comparison of internals of failing and valid executions can be used to explain why programs fail.
Instead of just comparing the coverage information, \emph{execution comparison} approaches compare traces and program states across executions~\cite{KKS+:ASPLOS15,WZSJ:ISSTA10, JCC+:SP11,Z:FSE02, SZ:ICSE13, SZ:FSE09, GCK+:STTT06, AOH:ASE07}.
Such comparison can highlight differences in data and control flow relevant to the fault.

Delta debugging is an approach that systematically narrows down the inputs (resp. changes) that are relevant for causing a fault~\cite{ZH:TSE02}.
Based on this idea, Zeller applied delta debugging to program states of executions~\cite{Z:FSE02}.
A challenge with delta debugging is that it needs to align statements and program states of independent executions~\cite{XSZ:PLDI08}.
Sumner et al.~\cite{SZ:ICSE13, SZ:FSE09} improved the initial work of Zeller using dual slicing~\cite{WZSJ:ISSTA10, JCC+:SP11} and efficient execution indexing~\cite{XSZ:PLDI08}.
They improve the efficiency and minimize the explanations in finding the cause effect chain~\cite{SZ:ICSE13, SZ:FSE09}.

Execution comparison approaches are designed to explain the differences between only two executions at a time.
Thus they require significant runtime overhead, as they execute the program many times to narrow down the instructions necessary for a certain behavior.
Due to the separate execution of the program, these approach need to deal with three major challenges, correct alignment of the executions~\cite{KKS+:ASPLOS16, XSZ:PLDI08, SZ:FSE10}, memory overhead that comes with recording the executions~\cite{KC:ISPASS11, BGJ+:TC05, KM:ICSE08}, and differences in executions caused by nondeterminism~\cite{KKS+:ASPLOS16}. \looseness=-1

\subsection{Understanding Feature Interactions}
\noindent
Feature interaction bugs are hard to detect as they are only triggered for certain combinations of features. 
There is a lot of research to efficiently find such faults, such as combinatorial interaction testing~\cite{CDS:ISSTA07, MKRGA:ICSE16,NL:CSUR11}, systematic sampling~\cite{KMS+13, SAG:ICSE17, useKBK:AOSD11}, model checking~\cite{RARF:JPF11, CHSL:ICSE11, BCMDH:LICS90}, and variational execution~\cite{NKN:ICSE14, MWK+:ASE16, KKB:ISSRE12,AF12}.

Sampling approaches can only reveal configurations that fail, but not the interaction that causes it.
Variational execution tracks the exact combination of inputs that lead to a fault, but does not help to understand \emph{why} the interaction happens and \emph{how} it causes the fault.\looseness=-1

Other approaches statically reason about the code to detect feature interactions.
The work of Kim et al. reasons about the combinatorics to reduce the number of configurations to execute~\cite{KBK:AOSD11}.
However, after testing these reduced configuration, the approach cannot answer \emph{why} configurations fail.
LoTrack reasons about which lines of code are affected by load-time options~\cite{LKB:ASE14}, but it cannot answer \emph{why} and \emph{how} options interact.
The research of Zhang and Ernst aims to identify configuration faults using thin slicing and a lightweight form of execution comparison~\cite{ZE:ICSE13, ZE:ICSE14, SFB:PLDI07}.
Their approach suggests single configuration options that are likely to trigger a fault. 
However, they assume that (a) the program itself is correct and (b) that a single option triggers the fault rather than a combination of multiple options.
\looseness=-1

In summary, there exist many approaches that help to compare two executions and approaches that help detecting interactions in larger configuration spaces. 
However, none of the existing approaches helps to understand \emph{how multiple options interact} as they either do not scale to multiple options or miss control and data-flow information.
In our work, we provide support that scales to explain interactions among multiple options and that also provides necessary control and data-flow information about the interactions.\looseness=-1

\section{Generating and Visualizing Variational Traces}
\label{sec:vt}
\noindent
In this section, we introduce the new concept of \emph{variational traces} which help developers understanding how options interact during the program execution and assist with debugging interaction faults.

\subsection{Variational Traces}
\noindent
We introduce a \emph{variational trace} as a compact representation of differences among multiple execution traces regarding control flow and the program states.
With variational traces, we explain how differences in inputs affect a program's execution and data, and how inputs interact with each other.
A variational trace is a graph that represents differences on the control-flow and on data using the following concepts (illustrated in Figure~\ref{fig:trace} for the program in Figure~\ref{fig:code}):
\begin{compactitem}[-]
\item \emph{State changes} (orange rectangles) describe statements that change values of fields and local variables.
In the example, the value of \emph{err} is changed from \emph{null} to -1 if the option \emph{flag} is true.\looseness=-1
\item \emph{Decisions} (diamonds) describe statements causing control-flow differences due to differences in inputs (directly or indirectly) between the individual traces (e.g., if-statements).
\item \emph{Decision Parameters} (gray rectangles) describe variables that are used in decisions.
In the example, \emph{err} is used as parameter for the decision of Line 12.
We found this information on parameters used in decisions particularly useful as they help to understand causes of control-flow differences.
\item \emph{Exceptions} (red rectangles) describe statements that are thrown due to faults or exception handling. 
In the figure, we see that the \emph{NullPointerException} is thrown under the condition that \emph{flag} is false and \emph{ipv4} is true.
\item \emph{Return} statements (not shown in the example) describe values that are returned by a method.
Return statements are included if a method returns different values or from different locations.
\item \emph{Methods} (rectangles around subgraphs) structure the variational trace and describe the stack trace (e.g., method \mbox{\texttt{getErrValue}}).
The notation of methods helps to understand the control flow and the execution of the program across method boundaries.
\end{compactitem}

\paragraph{Variational traces show differences} 
To be a useful debugging tool, variational traces need to concisely describe differences among executions while still containing sufficient information.
A variational trace only describes state changes, decisions, invocations and exceptions that differ among executions. 
Thus, a variational trace contains only statements that cause control-flow differences and statements that change variables differently in different executions.
Statements that do not cause such differences are not as important and will not be contained in the variational trace.
For example, the statement in Line 3 of Figure~\ref{fig:code} sets the value of \emph{err} to \texttt{null} in all configurations.
This state change is thus not relevant to describe differences among executions. 
The same applies to changes that only take place on variables that only exist for certain inputs.
Thus, if an object only exists under condition \emph{a}, then changes that happen to this object under condition \emph{a} are not important for comparing traces.
Instead, the variational trace will report changes that create interactions, such as Line 5 of the example that changes \emph{err} to be either -1 or \texttt{null} depending on the value of \emph{flag}.
Such a statement reduction requires to record and keep all state changes of all variables and to compare their values across all executions.\looseness=-1

A focus on differences among executions allows to narrow down the number of statements that are reported to the user.
Our approach might be further combined with slicing to remove statements that do not matter to explain a certain exception~\cite{Z:FSE02, SZ:ICSE13, SZ:FSE09}. 
In this work, we focus on efficiently generating explanations for differences among many executions.
Thus, improvements such as data-flow and impact analyses are out of this paper's scope.

A variational trace, as in Figure~\ref{fig:trace}, looks similar to a control flow graph.
However, three differences make variational traces more useful for debugging purposes:
(i)~variational traces represent actual executions and they show the actual values that variables take plus the exact context they are executed in,
(ii)~variational traces only contain statements that differ among the executions, and (iii)~variational traces support (beyond others) loops, recursion, dynamic invocation, and reflection.\looseness=-1

\begin{sloppypar}
\paragraph{Using Variational Traces} Variational traces help understanding differences among executions as they describe the cause and effects of differences among executions.
This allows answering questions about interactions among options and to understand faults caused by such interactions.
First, a variational trace answers for which inputs the interactions occurs.
This information is visible in the context of the statement (resp. exception).
Second, it helps understanding why the interaction occurs as it shows the differences in the state and in the control-flow at the point in the execution before the statement.
Third, a variational trace helps understanding how the interaction is caused and how the corresponding state is created.
As an interaction is always caused due to previous control-flow decisions or other interactions, the information about the cause will always be contained in the variational trace. 
For example, the exception in Figure~\ref{fig:formal} is thrown because the value of \emph{err} is null if \emph{flag} is false.
The cause of the fault is that \emph{err} was not initialized if \emph{flag} is false, which is visible in the trace.\looseness=-1
\end{sloppypar}

\subsection{Generating Variational Traces by \\Aligning Trace Logs}
\label{sec:formal}
\noindent
To generate variational traces we could align the traces from executing the traces from executing all configurations separately.
However, this way of creating variational traces would be challenging.
In this subsection, we use such a base line approach that aligns trace logs to explain the necessary steps and challenges of creating variational traces.
We illustrate each step with a running example in Figure~\ref{fig:formal}, resulting in the variational trace shown in Figure~\ref{fig:trace}.

The first step for creating a variational trace is to \emph{record} the program's executions for \emph{all} different inputs.
The recording needs to track both the instructions and the state changes, to understand the effects of the executions.
The recordings including state changes for the four inputs of the example are illustrated in Figure~\ref{fig:record}.

To compare the executions, all traces need to be aligned after recording.
To \emph{align} the traces, instructions from each trace that belong to each other need to be identified~\cite{NW:JMB70}.
In general, there are multiple different alignments possible as statements may repeat in different parts of the execution.
Thus, a semantically optimal alignment is required for meaningful results.
However, as the program is executed separately multiple times, alignment is a non-trivial task, especially as object references are different among executions, as it is necessary to keep track of iterations, and as nondeterminism may lead to wrong alignments, which can be avoided by recording and replaying nondeterministic result (e.g., IO) among the executions~\cite{KKS+:ASPLOS16, XSZ:PLDI08, SZ:FSE10}.
We illustrate the alignment of the traces in Figure~\ref{fig:align}.\looseness=-1

After aligning the traces, shared parts of the executions can be identified.
Thus, we can \emph{merge} the traces into a single trace with conditional statements~\cite{RC:FSE13}.
Statements and state changes that are included in several traces can be shared.
We illustrate merging of the four executions  from the example in Figure~\ref{fig:merge}.
As shown, Line 5 is only executed if the option \emph{flag} it true, and the value of \emph{err} depends on the input \emph{flag}.

Finally, the variational trace can be \emph{reduced} to statements that describe differences and can be enriched with information about the effects of the instructions.
We illustrate the variational trace for the example program in Figure~\ref{fig:trace}.
As shown, the variational trace only describes the control-flow and state differences among the executions resulting in a concise explanation of the differences and the fault.\looseness=-1

\begin{figure*}[t]
	\includegraphics[width=1\textwidth]{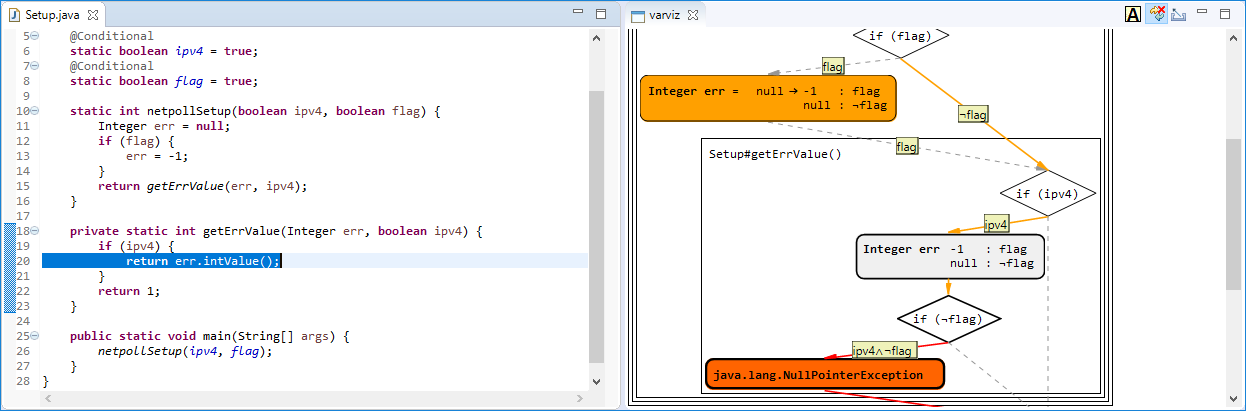}
	\caption{Screenshot of Varviz in Eclipse for the example of Figure~\ref{fig:code}: Java editor (left), variational trace (right).}
	\vspace{-8pt}
	\label{fig:screenshot}
\end{figure*}

\paragraph{Discussion} The baseline approach has the obvious issue that it can only scale to few options as it needs to run the program $2{^n}$ times for $n$ boolean options.
Furthermore, as it is unclear which statements will interact upfront, the approach needs to record full traces which causes severe memory problems. 
That is, the approach needs to keep all statements and states in memory until all configurations are executed.
There are several ways to reduce the memory consumption, such as 
(i) directly merging the traces after execution instead of keeping them all in memory, 
(ii) statically deciding which parts of a method will be equal independent of options (e.g., initialization of local variables),
(iii) compressed storage of trace logs~\cite{KC:ISPASS11, BGJ+:TC05}, and
(iv) inter-procedural analysis to detect statements that depend on options~\cite{LKB:ASE14}.
However, these approaches are either not sufficient (i, ii~\&~iii) or do not scale to larger programs (iv).\looseness=-1

We implemented the baseline approach in a prototype for Java including optimizations (i) and (ii).
We observed that the additional logging statements and the merging process cause a lot of computational overhead.
However, especially the memory consumption of the approach is problematic.
If too many statements and states must be kept in memory the tool may record multiple GB for the systems in our evaluation, as we show in Table~\ref{tab:programs2}, which makes this approach infeasible for larger systems.
To approach these challenges requires sophisticated mechanisms to reduce the number of statements to record and to solve the problems of nondeterminism (e.g., by synchronously executing all configurations~\cite{KKS+:ASPLOS15, KKS+:ASPLOS16}).\looseness=-1

Our goal is to generate variational traces anyhow as they enable us to observe effects and interactions of options.
To this end, we have to avoid all the limitations (scalability, memory overhead, nondeterminism) of the base line approach while keeping the idea of  aligning the executions of all configurations.\looseness=-1

\subsection{Efficient Generation of Variational Traces with Variational Execution}
\label{sec:varex}
\noindent
A key insight of our work is to use \emph{variational execution} to sidestep the previously discussed challenges for generating variational traces (i.e., handling the challenges of recording, alignment, merging and nondeterminism).
Variational execution is an approach that can efficiently execute \emph{all} configurations of the program in a single run~\cite{NKN:ICSE14, MWK+:ASE16}.
Scalability is achieved because all redundant parts among the executions are executed only once and redundancies in data are stored only once~\cite{MWK+:ASE16}.
If parts of the program are only executed under certain conditions, then also variational execution will run these parts only under these conditions.
For example, Line 5 of Figure~\ref{fig:code} is only executed if the option \emph{flag} is true.
Variational execution runs the program with all inputs as illustrated in Figure~\ref{fig:merge}.
Instead of executing parts multiple times as in Figure~\ref{fig:record}, each instruction is only executed once.\looseness=-1

To represent state differences between configurations, variational execution uses choice values~\cite{EW11}.
A choice value is a mapping from partial configurations spaces to concrete values.
For example, the value of \emph{err} in Figure~\ref{fig:trace} is stored as a choice that represents the state of \emph{err} for all four configurations.
Using a variational data representation allows variational execution to efficiently represent the states of all configurations as most data is redundant among configurations~\cite{MWK+:ASE16, M14}.
Sharing of redundancies in executions and data allows variational execution to scale for programs with huge configuration spaces that would not be practical with a brute-force approach that runs each input individually~\cite{MWK+:ASE16,NKN:ICSE14}.
We solve the problem of non-deterministic behavior among the configurations due to the sharing and synchronous execution of variational execution, for which an alignment-based approach requires sophisticated synchronization and alignment strategies~\cite{KKS+:ASPLOS16}.
\looseness=-1

\paragraph{Generating Variational Traces}
As variational execution synchronizes the executions among configurations, there is no need to align the executions of the program.
Instead, we directly generate the variational trace by recording how variational execution runs the program.
The context of each statement is already given as all instructions are executed under a certain context.
To observe where the execution splits, we just need to observe changes in the execution context.
Finally, we can observe interactions on data in the assigned choice values of local variables and fields.
With variational execution we avoid the memory explosion of the base line approach, as we can decide during execution whether a statement should be contained in the variational trace. 
As we will show as part of our evaluation in Figure~\ref{tab:programs2}, using variational execution is up to five times faster than the baseline approach while requiring up to 74\% less memory.
Beyond that, we also show that we can generate variational traces for huge configuration spaces for which it is simply impossible to generate them using a brute force approach (see Checkstyle).
\looseness=-1

To generate variational traces, we extended VarexJ, a variational interpreter for Java~\cite{M14, MWK+:ASE16}.
VarexJ is a metacircular interpreter that executes Java bytecode instructions conditional and that stores all data using choice values.
We adapted the execution of bytecode instructions, such as fields and local variable instructions to record their changes on data, if-statements to record whether they split the execution, method invocations to record the stack trace, and exceptions to report faults.\looseness =-1

While variational execution has been introduces before~\cite{NKN:ICSE14, M14}, our new contribution is to realize its potential for generating variational traces without the disadvantages (e.g., memory overhead) of aligning single traces, by observing internals of the variational execution engine.
Thus, with variational traces, we are usually able to align an exponential number of traces, which is practically impossible without it.

\paragraph{Generation with Symbolic Execution}
Variational execution shares similar ideas with \emph{symbolic execution}~\cite{C76, K76}.
Indeed, with symbolic execution it is possible to explore the executions for many inputs.
In contrast, to symbolic execution, variational execution always processes alternative but concrete values, not symbolic ones. 
Thus, variational execution does not have the problems of symbolic execution, such as decidable loop bounds.
As variational execution requires concrete inputs, our approach also requires a test case, which is given by our scenario as we want to compare executions for a given test.
Symbolic execution can hardly be used to generate variational traces as it typically does not share and align executions beyond common prefixes~\cite{CGK+:ICSE11, HHB:ASE16}.
MultiSE~\cite{SNG+:FSE}, which incorporates ideas from variational execution, introduces summary values to represent value differences, which enable MultiSE to increase the sharing abilities of symbolic execution. 
MultiSE could potentially be used to generate variational traces as the execution may be similar to variational execution and the data-flow differences can be observed in the summary values.
However, due to the overhead of the symbolic execution engine, it remains unclear whether MultiSE would achieve the same scalability and performance as variational execution.\looseness=-1

\subsection{Varviz}
\label{sec:varviz}
\noindent 
We argue that variational traces aid programmers to debug and understand programs by providing information about the program execution and interactions among options.
To make variational traces accessible, we implemented an Eclipse plug-in called \emph{Varviz} (from \emph{var}iation and \emph{vis}ualization).
We released Varviz as open-source \mbox{(\url{https://meinicke.github.io/varviz/}).}
The plug-in already comes with the utilized VarexJ to generate variational traces.\looseness=-1

In Figure~\ref{fig:screenshot}, we show a screenshot of Varviz.
The variational trace can be generated using default run mechanisms of Eclipse, which will automatically call VarexJ.
After running the program, the variational trace is shown in the Varviz view (shown on the right).\looseness=-1

Navigation is one of the most time-consuming tasks during debugging~\cite{KMC+:TSE06}.
Therefore, we designed Varviz to also be used as a navigation tool.
By double-clicking on elements in the trace, the tool automatically displays the file and the line of the element.
As shown in the screenshot, the return instruction that throws the exception is highlighted after double-clicking the exception statement in the trace.\looseness=-1

\paragraph{Focus on selected interactions}
In practice, many interaction faults occur among few options~\cite{MWK+:ASE16, \samplingevidenceforlowdegree}.
To understand a certain interaction among a specific set of options, we show a, usually smaller, relevant parts of the variational trace.
To this end, we provide a new \emph{projection mechanism} to show only the statements that explain the effects of a given set of options.
All other options are set to fixed values, false if possible (the valid selection might be restricted by a feature model~\cite{KCHNP90}).
By setting all other options to fixed values, Varviz will produce a \emph{projection} for the interaction of the few options of interest.
For example, if a fault is thrown under context $A\wedge\neg{}B$, we are interested in the interactions of these two options, but not in options $C$ and $D$.
To create a projection on the variational trace for $A$ and $B$, we set the other options $C$ and $D$ to false (if possible) to hide the effects and interactions of these options.
To remove unnecessary elements, we create the constraint $\neg C \wedge \neg D$ and evaluate the conditions of all elements of the variational trace.
If the condition of the element under the context of the constraint is satisfiable, we keep the element in the variational trace, otherwise it can be removed.
Finally, we evaluate the remaining elements, whether they represent differences among the executions for the options of the projection. 
Removing options from the trace can highly reduce its size and thus helps to understand the interactions (as we will show in Section~\ref{sec:quantitativeanalysis}), while preserving interactions that are relevant for the options of interest.\looseness=-1

\subsection{Limitations of Variational Traces}
Variational traces inherit limitations from related automatic debugging techniques based on trace alignment~\cite{Z:FSE02, SZ:ICSE13}.
Similar to these techniques, we compare separate executions to explain causes of faults and interactions.
The similarity of these executions determines the quality of variational traces. 
For example, executions (i.e., test cases) can introduce minor changes (noise) that are irrelevant to the fault of interest.  
Thus, executions that minimize noise are always preferable.
In contrast, inputs need to trigger similar executions to reveal enough information about fault and its cause.
If the executions are too different, then the variational trace cannot provide enough information.
When we apply variational traces to configurable systems for the same test case, the executions of the configurations will be similar by design. 
However, if a fault is not caused by an interaction, but simply because certain code is executed, then we can only report the context and location of the fault but not necessarily its cause.
\looseness=-1

We use the state-of-the-art variational execution engine VarexJ to generate variational traces~\cite{M14, MWK+:ASE16}. 
However, VarexJ has engineering limitations inherited from the interpreter of JavaPathfinder~\cite{HP:STTT00}, such as incomplete support for native methods~\cite{SB:SPIN14}, multi-threading, and performance overhead due to interpreting Java bytecode.
Overall, VarexJ is mature enough, including complex language features such as reflection, to be able to execute several medium sized Java programs~\cite{MWK+:ASE16}, but not larger industrial sized programs due to the runtime overhead and its technical limitations.
Variational execution is an evolving technique and advancements in variational execution will also improve the efficiency and applicability of our approach.\looseness=-1

\newcommand{\RQ}[2]{\textbf{RQ#1:} \emph{#2}}

\section{User Study}
\label{sec:userstudy}
\noindent
Automated debugging techniques often promise large effects for debugging tasks.
Previous evaluations on approaches based on execution comparison focused on reporting the size of the explanations (number of statements) instead of showing weather and how helpful they actually are for debugging~\cite{SZ:ICSE13, Z:FSE02}.
However, only reporting quantitative results of the approach can be misleading  and the expectations may not meet the reality (e.g., the explanations may be too complex and complicated to be understandable or do not contain the necessary information to understand the fault)~\cite{PO:ISSTA11}.
This is the first user study on delta debugging like approaches that we are aware of.

We designed variational traces to help users to understand variations in executions.
In our evaluation, we investigate how and why variational traces help users.
Specifically, we perform a user study to answer the following research questions:\\[0.3em]
\RQ{1}{How much do variational traces improve the performance of solving debugging tasks compared to a standard debugger?}
To answering RQ1, we explore the speedup and the success rates for solving debugging tasks.
\\[0.3em]
\RQ{2}{How do variational traces help understanding differences in executions?}
With RQ2, we investigate what the information needs are during a debugging task and whether the variational trace can answer them.
We want to evaluate whether the provided information (i.e, the statements shown in the variational trace) is sufficient to help understanding the interactions.\looseness=-1

\paragraph{Systems and Tasks}
\noindent
We use three subject systems in our evaluation, namely GameScreen, Elevator and NanoXML.
Statistics on the programs are shown in Figure~\ref{tab:programs}.
We carefully chose these systems for different reasons:

\begin{figure}
	\centering
	
	\resizebox{\columnwidth}{!}{\begin{tabular}{ l r r r r r r r r}
	\\[-6pt]\toprule
		Program & LOC & Cov & Opt & Conf & M & N & D & Instr\\
		\midrule
		GameScreen & 32 & 32 & 3 & 8 & 4 & 12 & 6 & 230\\
		Elevator & 259 & 193 & 6 & 20 & 7 & 12 & 3 & 5,688\\
		NanoXML & 1000 & 331 & 1 & 2 & 18 & 21 & 4 & 42,138\\
		\bottomrule\\[-1.5em]
	\end{tabular}
}
	\caption{Statistics on the programs used in the user study (Cov: Covered LOC by the test case, Instr: instructions executed for the  test case,  M: Methods, N: Nodes, D: Decisions).}
	\label{tab:programs}	
\end{figure}

\emph{GameScreen} was previously used in a study which conducted the effect of different degrees of variability on program comprehension~\cite{MBW:ICSE16}.
The program is a code snippet inspired by a real variability bug in BestLap, a configurable race game.
Melo et al. have shown that the small program with only three features takes on average ten minutes to debug without tool support~\cite{MBW:ICSE16}.
The program contains a fault that is triggered by the interaction of two features. 
The task is to understand the cause of the fault and the configuration in which the fault appears.
The program is too trivial and cannot give any insights for our study as the program can be understood in few minutes using a standard debugger. 
Thus, we used GameScreen only as warm-up task to make the users familiar with the type of tasks they should perform.\looseness=-1

\emph{Elevator} is a simulation of a configurable elevator system~\cite{PR01}.
The program is designed to trigger interactions among its options.
Even though the program has only few lines of code, it is hard to understand the impact of its features due to the interactions.
The program comes with several specifications in form of runtime assertions that are violated for certain configurations.
We selected a specification that states that the elevator should continue in its current direction if there are still calls in this direction.
This specification is violated if a feature for \emph{executive floors} is on, which can force the elevator to change its direction.
In the tasks for Elevator, the participants should figure out in which configurations the fault appears and how the fault is caused.
Although fixing a fault is part of debugging, fixing itself is not part of the task as this would have required to change the program's specification.
Instead it was sufficient to explain how the fault is caused.\looseness=-1

\emph{NanoXML} itself is not a configurable system.
The program was used in a prior study to evaluate whether the automatic debugging technique Tarantula can help programmers with debugging~\cite{PO:ISSTA11}.
Tarantula showed only minor improvements for debugging NanoXML compared to a standard debugger.
We evaluate NanoXML on the same bug as in the original study, in line with the original study~\cite{PO:ISSTA11}.
We provide two slightly different files as input for parsing.
One of the files cannot be parsed correctly, causing an exception. 
The other one is a similar file that can be parsed.
Both files are parsed simultaneously using variational execution.
In addition to the tasks of the previous programs, the participants were also asked to fix the bug similar to the prior study~\cite{PO:ISSTA11}.
With NanoXML we show that variational traces are helpful for a standard debugging tasks to understand variations beyond configurable systems.
Thus, with the NanoXML experiment we can show the usefulness of comparative- and delta-debugging approaches which have not been evaluated in user studies before~\cite{SZ:ICSE13, Z:FSE02}.
\looseness=-1

\paragraph{Pilot Study}
We performed a pilot study to estimate the required power (i.e., number of participants) of our study and to tune the task and descriptions.
We asked several graduate students to use Varviz and the Eclipse debugger on our tasks.
We found and revised several issues of the usability of Varviz.
We also measured the time and estimated that the effect size was big enough to show significant effects with few participants in the actual experiment.
For Elevator we had an estimation of 40 minutes when using the Eclipse debugger, compared to an estimation of 12 minutes when using Varviz.
For, NanoXML we have an estimation of 22 minutes from a previous study when using a debugger, which we use as estimation when using a standard debugger (our results were slightly higher)~\cite{PO:ISSTA11}.
We have an estimation based on our pilot study of only 8 minutes when using Varviz.
\looseness=-1

\paragraph{Study Design}\noindent
We designed our experiment as between-subject study to compare performances between participants using a standard debugger (baseline) or Varviz (treatment).
We did not mix the participants between using the standard debugger and Varviz (i.e., within-subject design). 
Each participants solved all three tasks with the same tool to reduce training time required for Varviz and to avoid carryover effects, such as learning effects and demand effects~\cite{CGK12}.
Learning effects from the first tasks might be applied to the second which influences the performances when using different tools.
Also, the motivation of using a new tool can influence the performance of the participants.
This effect is amplified if they are using both tools in a within-subject design~\cite{CGK12}.
A between-subject design has less statistical power compared to a within-subject design (i.e., we may need more participants to show significant effects), however, as we expect the effect size to be very large, as suggested by the literature on user studies~\cite{CGK12}, a between-subject design is more appropriate as it avoids confounding factors of within-subject designs.
\looseness=-1

We did not design two comparable tasks, but intentionally two very different ones for external validity. 
A within subject design typically requires multiple similar tasks, which is a benefit using a between-subject design.
While the participants worked on the tasks, with their consent, we recorded the screen and asked them to verbalize their thoughts using think-aloud protocols~\cite{BH:97}.
These recordings help us to track the participants' information needs and debugging strategies.

Other approaches, such as delta debugging~\cite{Z:FSE02} and comparative causality~\cite{SZ:ICSE13} may give similar textual explanations of the faults.
However, we cannot compare our approach with delta debugging~\cite{Z:FSE02} and comparative causality~\cite{SZ:ICSE13} as the tools are not available (we contacted the authors) and as they are designed to explain differences among only two executions.\looseness=-1

\paragraph{Methods}
To answer RQ 1, we compare the time and success rates of the participants for solving the tasks.
To answer RQ 2, we record the audio and the screen of the participants. 
We analyze the recordings based on qualitative content analysis using open coding~\cite{S:12, S15}.
We watch the videos with the goal to find common tasks that the participants perform during debugging.
We use these commonalities to create a coding frame that allows us to understand how the participants perform when using Varviz or the standard debugger.\looseness=-1

\paragraph{Participants}\noindent
As we plan to perform think-aloud protocols, analyzing the data (i.e., screen recordings and audio) requires high effort.
We thus aim to avoid an unnecessary high number of participants.
According to Nielsen~\cite{N:IJHCS94}, for think aloud protocols five participants are sufficient to gain most insights. 
Adding more participants does not give more essential information.
Thus, to answer RQ 2, we require at least five participants per group, so ten participants in total.\looseness=-1

To answer RQ 1, we calculate the minimum number of number of participants required using power statistics based on the pre-study results.
We use the Rosner's equation to calculate the required sample size $n$ for each group in our study~\cite{R15}:
\begin{equation}\label{eq:1}
n = \dfrac{(\sigma_1^2 + \sigma_2^2) (z_{1-\alpha /2} + z_{1-\beta})^2}{\Delta^2}
\end{equation}

\noindent We use a conservative $\alpha$ value of 0.01 which is a probability of Type \Romannum{1} error of 1\% (i.e. the probability to find an effect even though there is none, usually 0.05).
We use a conservative $\beta$ value of 0.05 which is a probability of Type \Romannum{2} error of 5\% (i.e. the probability not finding an effect even though there is one, usually 0.2).
The statistical power is $1 - \beta$ and thus 95\%. 
We use a high value for the estimated standard deviations $\sigma_1$ and $\sigma_2$ of 5 minutes as we only used few participants in our pilot study.
Using Equation \ref{eq:1} and our estimations of expected performances (10 versus 40 and 8 versus 22 minutes for Elevator and NanoXML respectively), we can calculate the number of participants needed for our experiment as one participant ($n_{elevator} = 0.989 \approx 1$) and five participants ($n_{NanoXML} = 4.544 \approx 5$) per group for Elevator and NanoXML respectively.
As the required group size for NanoXML is larger than for Elevator, we use a group size of five for our experiment.
Thus, based on our pre-study results, we need ten participants to show significant results.
For both, our quantitative and our qualitative analysis ten participants are sufficient.
\looseness=-1

We recruited ten participants at Carnegie Mellon University: eight undergraduate students, one graduate student and one post doc.
The participants were recruited using posters and mailing lists.
We excluded participants without prior knowledge of Java.
All participants received a 25-dollar gift card after finishing the experiment.
The participants were assigned randomly to two groups of five.
One group used the Eclipse debugger, the other group used Varviz.
The graduate student used Varviz, while the post doc used the Eclipse debugger.\looseness=-1

Before conducting the experiment, we asked the participants for their programming experience and experience with Java.
The groups were roughly similar: 
The median experience for programming is 3.5 years for both groups (average is 3.7 years for the debugger group and  5.3 years for the Varviz group;
note that the large difference in average experience is caused by as a single outlier (Varviz 1) who reported 12 years of non-professional programming experience.)
The median Java experience is 2 years for both groups (average is 2.6 years for the debugger group and 2 for the Varviz group).
None of the participants knew variational execution or any of the subject programs.
All participants in the debugger group have used Eclipse and the debugger before.
If a participant did not remember how to get to a certain view, such as the call hierarchy, or were unsure about certain functionalities of Eclipse, we provided this information during the experiment.

\paragraph{Execution}\noindent
Both groups were given an Eclipse containing the three programs they had to debug including a failing configuration for Elevator and the two XML files for NanoXML.
The participants using Varviz were introduced to the functionalities of the tool.
We used a simple foo-bar example to explain the functionalities of Varviz.
All participant started debugging the programs in the same order, first GameScreen, second Elevator and third NanoXML.
The participants performed the tasks for Elevator and NanoXML until they solved them correctly, until they gave up, until they reached a time limit of 30 minutes per task, as we planned the experiment to take roughly one hour.\looseness=-1

As we performed a think-aloud protocol, we conducted the experiment with each participant in isolation.
To record the audio and screen, we used an inhouse recording tool from our university.
For the first participant using the debugger (Debugger 1), we used a different open source recording tool.
We lost the screen recording due to a fault in the recoding tool. 
However, we kept the results and the audio recording from this participant in our study as he had a good performance compared to the others using the debugger.

\begin{figure*}[t]
	\centering
	\begin{subfigure}[b]{0.98\textwidth}
		\includegraphics[trim=00pt 300pt 0pt 0pt,clip,width=1\textwidth,page=7]{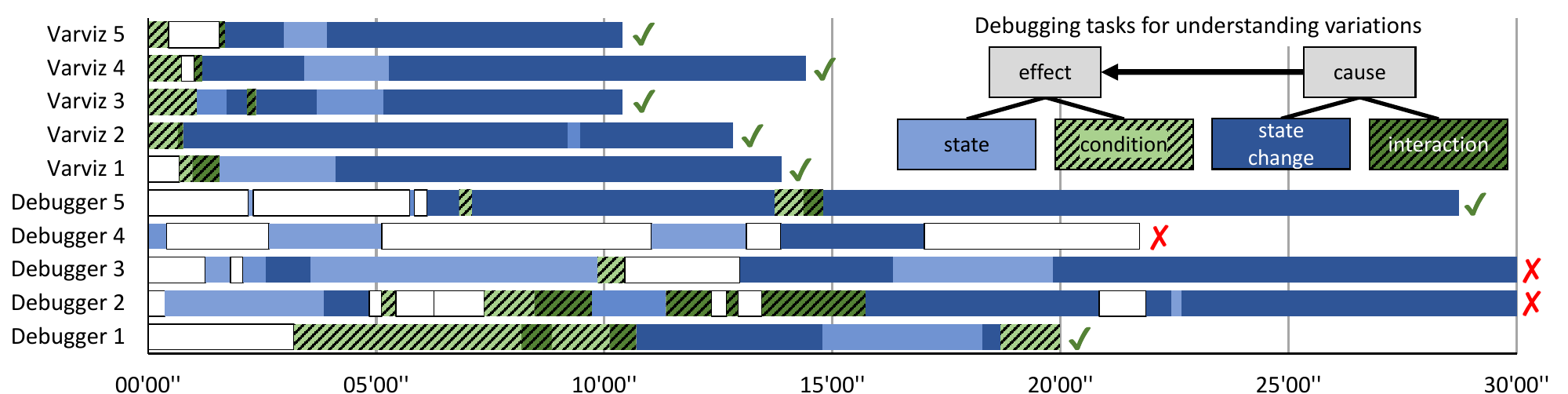}
		\caption{Elevator}
		\label{fig:timeline:elevator}
	\end{subfigure}
	\begin{subfigure}[b]{0.98\textwidth}
		\includegraphics[trim=0pt 300pt 0pt 0pt,clip,width=1\textwidth,page=6]{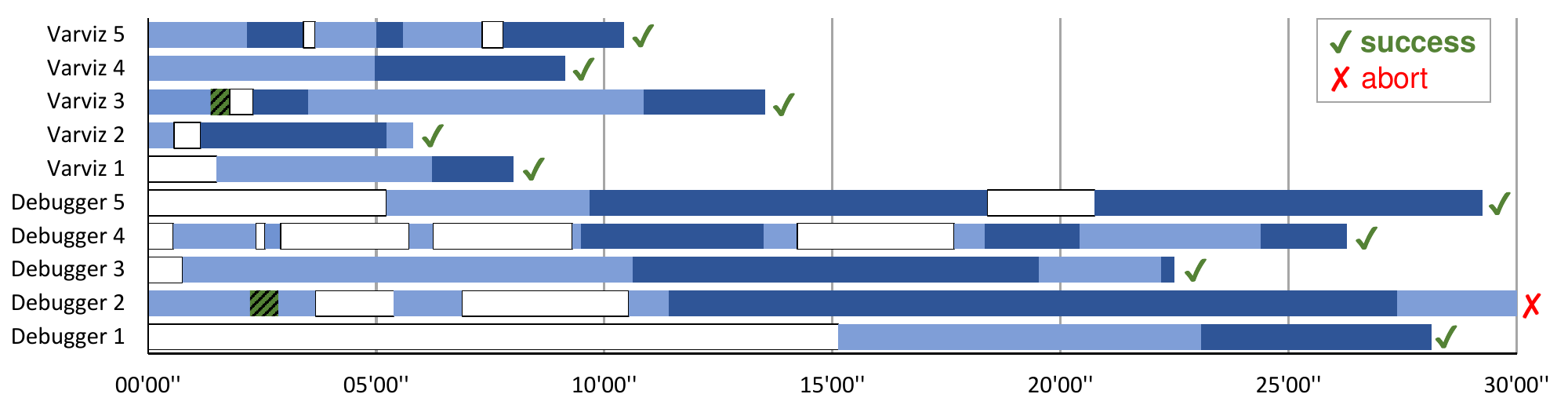}
		\caption{NanoXML\\[-0.5em]}
		\label{fig:timeline:nanoxml}
	\end{subfigure}
	\caption{Time spend on debugging tasks. White boxes denote unrelated tasks.}
	\label{fig:timeline}
\end{figure*}

\definecolor{mygreen}{RGB}{100,123,84}
\definecolor{myred}{RGB}{185,58,58}

\paragraph{Quantitative Analysis}
\noindent
To answer RQ1, we compare the time and success rates of the participants for solving the tasks.
In Figure~\ref{fig:timeline}, we plot the performances of the participants.
The signs {\color{mygreen}\textbf{\ding{51}}} and {\color{myred}\textbf{\ding{55}}} indicate correct and aborted solutions respectively.
When using the Eclipse debugger only two out of five participants solved the Elevator task correctly and four out of five for NanoXML.
In contrast, all participants using Varviz solved the tasks for both programs.

For Elevator the participants using Varviz took on average 12 minutes while the participants using the debugger took on average 28 minutes.
The best performance using the debugger took 20 minutes, more than five minutes longer than the worst participant using Varviz.
For NanoXML we see similar results.
The Varviz group took on average 9 minutes.
In contrast, the debugger group took on average 27 minutes (best was 22 minutes).
Our results are in line with prior research which reported an average time of 22:30 minutes using only a debugger~\cite{PO:ISSTA11}.
To calculate the \emph{effect size} of using Varviz compared to a standard Debugger, we use the corrected equation of \emph{Hedges' g} which corrects for the upwards bias on small sample sizes:
\begin{equation}\label{eq:g}
g = \dfrac{\mu_1 - \mu_2}{\sqrt{\dfrac{\sigma_1^2 + \sigma_2^2}{2}}}\times\dfrac{N - 3}{N - 2.25}\times \sqrt{\dfrac{N-2}{N}}
\end{equation}
\noindent A $g$ value of 1 indicates that the groups differ by 1 standard deviation. 
In general, a $g$ value larger than 0.8 indicates a large effect size.
We calculated a $g$ values of 3.6 for Elevator and 4.9 for NanoXML.
Thus, $g$ values indicate a huge effect of using Varviz compared to a standard Debugger for understanding differences among executions. 
The differences are statistically significant (Mann-Whitney U test: p < 0.01).
\looseness=-1

\begin{framed}
\noindent\RQ{1}{How much do variational traces improve the performance of solving debugging tasks compared to a standard debugger?\looseness=-1}\\[0.5em]
The participants using Varviz are on average \textbf{55.4\%}, respectively \textbf{65.5\%} faster than participants using the Eclipse debugger for Elevator respectively NanoXML.
The success rates when using the Eclipse debugger are \textbf{40\%} for Elevator and \textbf{80\%} for NanoXML.
When using Varviz, the success rates are \textbf{100\%} for both programs.
\end{framed}

\begin{sloppypar}
\paragraph{Qualitative Analysis}\noindent
To answer RQ 2, we analyzed the audio and the screen recordings to understand how variational traces help understanding differences in executions.
Our coding frame~\cite{S:12, S15} is shown in the right upper corner of Figure~\ref{fig:timeline:elevator}.
Overall, all participants try to understand the relationship between \emph{effect} (e.g.,~the fault) and its \emph{cause}~\cite{LGHM:FSE07}.
The tasks for understanding effect and cause can be refined further.\looseness=-1
\end{sloppypar}

To understand the effect, it is necessary to understand program state and the condition.
The program state is necessary to know values of variables and which method calls are important.
When analyzing variations, it is also important to know under which condition (i.e., selection of options) the fault appears.

After understanding the effect, it is possible to investigate its cause.
It is necessary to understand the state changes and method calls that lead to the state of the effect.
As the effect only happens under certain conditions it is also necessary to investigate how specific selections of options cause the effect. 

In Figure~\ref{fig:timeline}, we used our coding frame to illustrate on which tasks the participants were working on.
Additionally, to the tasks in our coding frame, the participants also spend time with unrelated tasks, such as reading, scrolling, or investigating unrelated code~\cite{KMC+:TSE06}.
Unrelated tasks are shown with white boxes.
We plot the tasks on a horizontal time axis.
The group using Varviz performed the tasks much better than the group using the debugger.
In the following, we investigate the reasons why the tasks are so difficult when using only a debugger.
We explore which information are required to solve the tasks and how Varviz helps to gather them.\looseness=-1

We can see in Figure~\ref{fig:timeline} that the debugger group spend much more time on unrelated parts of the program and on tasks that do not lead to solving the problem. 
The main reason is that the participants read unrelated code of the programs.
Another reason is that the participants give up on their current goal and try to get information from other places.
When using the debugger, it is up to the programmer to find the places where to find information about the program.
Thus, the participants were lost in the source code they did not know, which leads to confusion and reading of code.
In contrast, when using Varviz, the participants had a guide that helps them to find the few locations in the code that are of interest.

When we analyze the performances for Elevator more closely (see Figure~\ref{fig:timeline:elevator}), we see that the Varviz group took only little time to understand the variations.
All the participants almost instantly figured out the condition of the fault (which is trivial using the variational trace as it is indicated by the context of the exception statement).
Also, figuring out where the option affects the behavior of the program is simple using Varviz as there are only few place (few decisions) where the option affects the data flow.
In contrast, when using the debugger, the first task is to figure out the condition of the fault. 
Without specialized tool support, this requires switching the options in the configuration and to re-execute the program. 
The time spent on this task depends on how many options the program has, how many options interact, and finally on luck or intuition as for participant \emph{Debugger 5}.
A simple tool that reports the condition of the faults (e.g., brute force) would help the debugger users and would improve their performances.
However, such a tool alone is not sufficient as it solves only a small part of the problem.\looseness=-1

After finding the condition of the fault, the participants still need to answer how the option triggers the effect.
By searching where the option is used, the place can be found, however also unrelated usages and only direct usages are found.
Thus, participants \emph{Debugger~3} and \emph{Debugger~4} did not even identify this part of the program.\looseness=-1

Finally, we can see that the Varviz group spend little time to understand the state at the exception.
This means that they can spend more time for understanding how the interaction triggers the effect and how this causes the fault.
The debugger group took overall more time to find the values of the variables and their values at the exception (except of participant \emph{Debugger 5} who performed well on this task).
The tasks to identify the exception state and the state changes are particularly hard as the program calls the scheduling method for the elevator in a loop.
This makes setting breakpoints hard as they are triggered multiple times before the actual state of interest.\looseness=-1

In the performances for NanoXML (see Figure~\ref{fig:timeline:nanoxml}), we can see that both groups struggle for identifying the state of the exception and answering why this state causes the fault.
This is because of a relatively difficult if-statement shown in the listing below.
The variational trace provides the values used in this if-statement.
However, the participants still need to understand the meaning of it.\looseness=-1

\begin{lstlisting}[escapechar=$, label={lst:nanoxml}]
if (!str.equals(prefix==null?name:prefix+name))
	XMLUtil.errorWrongClosingTag(this.reader, name, str);
\end{lstlisting}

\noindent After the participants using Varviz understood the meaning of the if-statement, they only spend little time to find the place of the cause as it is pointed out by a decision in the trace.
In contrast, the debugger group again struggled to identify the cause.
One reason is that the parsing is implemented using recursion, which again means that the breakpoints are triggered multiple times at the wrong state.
This again shows that control-flow (i.e., the decision of the cause) and data-flow information (i.e., the values that differ among the two executions) are essential to understand the differences among executions.
Both information are contained in the variational trace.

\begin{framed}
\noindent\RQ{2}{How do variational traces help understanding differences in executions?}\\[0.5em]
\noindent
Understanding where and how configuration options influence the execution is hard using only a standard debugger as only one configuration can be executed at a time. 
Thus, participants using the Eclipse debugger have difficulties figuring out the \textbf{condition} of the fault, gathering information about differences in the \textbf{program states}, and to find the \textbf{cause} of the fault.
Variational traces help with these task by providing essential information about the \textbf{fault condition}, the \textbf{fault state}, as well as \textbf{data and control-flow differences} that lead to the fault.
This information helps users to focus on important parts of the execution which additionally prevents from wasting time on unrelated activities.\looseness=-1

\end{framed}

\paragraph{Threats to Validity}
We designed our user study as between-subject study with only ten participants.
As discussed, we decided to use a between-subject design to avoid confounding factors and to reduce training time at the cost of the reduced power of the design~\cite{CGK12}.
We carefully calculated the required number of users upfront to minimize the effort for analyzing the think-aloud protocols and the screen recordings~\cite{R15}.
The measured performances in the experiment approximately match our expected performances form our pilot study.
Due to the large effect sizes a larger number of participants is unnecessary. 
The fact that our results are statistically significant confirms that the chance of a random error due to small sample size is small.

To reduce selection bias, we recruit participants in public channels and randomly assign them into two groups.
The average programming experience varies by almost two years between the groups, which is however caused by a single outlier (Varviz 1).
Our results are robust to removing this outlier (i.e., our results remain statistically significant without this participant)~\cite{LLT:APSEC14}.
To avoid effects due to differences in programming experiences, we designed the tasks in a way that basic debugging experience is sufficient.
Since we conduct the user study in the Eclipse environment, experience with the Eclipse toolchain is likely to affect performance of participants.
We performed a warm-up task familiarize the participants with the type of tasks and the programming environment.
We argue that most common usages of Eclipse are straightforward to most developers, given that Eclipse is a standard and classic development tool for Java.
In addition, we made it clear to the participants before study that we could provide immediate support for questions on Eclipse usage. 
However, participants rarely asked for help regarding Eclipse. 
Thus, we argue that Eclipse experience likely has only minor impact to the performance results.
To minimize confounding factors, we implement Varviz in a way that is completely orthogonal with existing features of Eclipse. Moreover, our introduction to Varviz only covers the plugin itself, not including any other functionalities of Eclipse.
The think-aloud protocol influences the time performance of the participants; however, it influences both groups equally and because the expected effect size is big we do not expect any systematic influences on the overall results.\looseness=-1

We used two diverse systems for the debugging tasks.
We showed that variational traces are useful to understand faults in systems with multiple options as well to compare two executions.
However, readers should be careful when generalizing our results to other systems and tasks.\looseness=-1

\section{Scalability Evaluation}
\label{sec:quantitativeanalysis}
\noindent
Variational traces are concise representations of differences among executions.
To further reduce the size, we allow focusing on small sets of options discussed in Section~\ref{sec:varviz}.
However, we do not yet apply any kind of impact analyses to reduce the size even further, as this is out of scope of this paper and as the sizes are already small enough especially for the programs used in Section~\ref{sec:userstudy}.
Variational traces are useful beyond debugging, as for example in our work on detect behavioral feature interactions with feature interaction graphs which can deal with large variational traces~\cite{SMN+:VaMoS18}.
In this section, we evaluate the size of variational traces when aligning the executions of exponentially large configuration spaces. 
Specifically, we answer the following research question:\\[-0.5em]

\noindent\RQ{3}{How does the generation of variational traces using variational executions scale compared to a base line approach?}
The exponential growth of configuration spaces with the number of options is challenging for both, execution and alignment of many configurations.
By answering RQ 3, we investigate the scalability of using variational execution to generate variational traces with regard to runtime and memory consumption.

\noindent\RQ{4}{How large do variational traces get?}
Information on data and control flow differences are beyond debugging (cf. Section~\ref{sec:application}).
With RQ4, we investigate how complex variational traces get when applied to programs with different numbers of options and different sizes.
We also want to find out how effective the filters for options are for reducing the size (see Section~\ref{sec:varviz}).

\paragraph{Experimental Setup}
In our evaluation, we reuse six configurable systems from our previous study on feature interactions and essential configuration complexity~\cite{MWK+:ASE16}.
The systems are shown in Figure~\ref{tab:programs}.
The systems are from different domains and show different interaction properties~\cite{MWK+:ASE16}.
We execute each system for a corresponding standard scenario.
Each system comes with a set of options that can be enabled and disabled, which results in large numbers of configurations for which we execute the program (see Figure~\ref{tab:programs2}).

\newcommand*\mean[1]{\bar{#1}}

\begin{figure*}
	\centering
\resizebox{\textwidth}{!}{\begin{tabular}{ l r r r r r r r r r r r r}
			\\[-6pt]\toprule
			Program & SLOC & Opt. & Conf. & Instructions & Time & TimeBL & Memory & MemoryBL & D$_{all}$ & D$_3$ & S$_{all}$ & S$_3$ \\
		\midrule
		CheckStyle & 14,950 & 141 & $>2^{135}$ & 194,725,919 & 209.8s & *364.8.8s & 1984MB & *3269MB & 5,989 & \textit{165.50} & 290,477 & \textit{2022.7}\\
		QuEval~\cite{SGS+:VLDB13} & 3,109& 23 & 940 & 6,460,571 & 10.6s & 50.3s & 379MB & 1498MB &699& 26.9 &4,152 & 110.0\\
		GPL~\cite{LB:GCSE01} & 662 & 15 & 146 & 17,457,437 & 15.6s & 18.9s & 408MB & 975MB & 530 & 6.9 &5,565 & 301.0\\
		Elevator~\cite{PR01} & 730 & 6 & 20 & 23,559 & 0.1s & 0.2s & 41MB & 29MB & 36 & 17.7 &96 & 51.0 \\
		E-Mail~\cite{H:ASE05} & 644 & 9 & 40 & 25,846 & 0.2s & 0.4s & 28MB & 49MB & 53 & 9.7 & 129 & 17.5 \\
		Mine Pump~\cite{KMSL:CDT83} & 296 & 6 & 64 & 21.615 & 0.1s & 0.4s & 37MB & 49MB & 10 & 7.6 &14 & 10.9 \\
		\bottomrule\\[-1.5em]
	\end{tabular}}
	\caption{Statistics on programs used in quantitative evaluation. D$_{all}$ and S$_{all}$ state the number of decisions respectively statements of the full variational trace. D$_3$ and S$_3$ state the mean number of decisions respectively statements after filtering for three options. In TimeBL and MemoryBL we show the time and memory consumptions for the baseline implementation (*for CheckStyle we only executed the configurations for five options with the baseline approach).}
	\label{tab:programs2}

\end{figure*}

The experiments are performed on a Windows computer with 8 GB ram and an Intel i5 processor with 4 cores.
To answer RQ 3 we collect the execution time and the memory consumption, we ran variational execution and the base line approach ten times and report the median value to avoid measurement errors.

The size of the variational trace depends on several factors.
First, the length of the execution (see number of instructions in Figure~\ref{tab:programs2}).
It also depends on how the program implements variability and how the options interact.
The fewer options interact in a program, the shorter the trace will be (statistics on the interactions in the systems are discussed by Meinicke~\cite{MWK+:ASE16}).
In our evaluation, we collect metrics on how large variational traces can get for different implementations and executions.\looseness=-1

Statements and decisions indicate the complexity of the variational trace.
However, understanding a fault or an interaction usually only requires few options as the interactions degrees are usually low~\cite{MWK+:ASE16, \samplingevidenceforlowdegree}.
We use context filters as discussed in Section~\ref{sec:varviz} to filter the variational trace for all combinations of three options (we filter the trace of CheckStyle only for \emph{one} option due to the large number of options).
\looseness=-1

\paragraph{Results}
In Figure~\ref{tab:programs2}, we report the times and memory consumptions required to generate the variational trace.
As shown, the time to generate the variational trace is always lower than with the base line approach, especially for larger configuration spaces.
For Checkstyle with 141 options, our approach takes 209.8 seconds, which is lower than what the base line approach takes when aligning the configurations for only five options.

As expected, the memory consumption of the base line approach becomes problematic, especially when aligning longer execution traces.
Again, the base line approach requires 3 GB for aligning only the traces for five options while our approach takes 2 GB when aligning the traces for \emph{all} configurations.
Note that the reported memory consumption of our approach also contains the memory overhead of VarexJ itself.\looseness=-1

\begin{framed}
	\noindent\RQ{3}{How does the generation of variational traces using variational executions scale compared to a base line approach?}\\[0.5em]
	\noindent Generating variational traces using variational executions scales to exponentially large configuration spaces with regard to execution time and memory consumption.
	In contrast, the base line approach is only able to generate variational traces for smaller configuration spaces while taking more time and memory than our approach based on variational execution.
\end{framed}

In Figure~\ref{tab:programs2}, we report the sizes of complete variational traces and the mean sizes after applying filters for three options.
Even though the interaction experiments (Elevator, E-Mail and Mine Pump) are designed to cause many interactions among options, we see that the sizes of their variational traces are small.
For GPL the complete variational trace becomes large as the executions contain long iterations which cause trivial but repetitive interactions on data (e.g., optionally initializing the weight for all vertexes in a graph). 
The same happens for QuEval which also has trivial but repetitive executions. 
For CheckStyle, which has by far the longest executions, we see that the size of the complete variational trace is huge.
The size of the variational trace is, however, small in relation to the configuration space and the number of executed instructions.
Most of these statements in the variational trace for CheckStyle are again repetitive.
After applying the filter for one option, we can see that the size can be reduced a lot.
In general, the sizes of the traces become small after filtering them. 
Even the full traces can be useful as most of the shown statements and decisions are repetitive due to iterations.\looseness=-1

\begin{framed}
	\noindent\RQ{4}{How large do variational traces get?}\\[0.5em]
	\noindent
	Variational traces are often small, but can become large, especially due to long iterations that repetitively create the same interactions on data.
	The size can be drastically reduced due to the filtering mechanisms, such as filtering for a small set of options.\looseness=-1
\end{framed}

\vspace{-.4em}
\paragraph{Threats to Validity}
To mitigate threats to external validity, we analyze different programs that show different kinds of interactions. 
We argue that variational traces are scalable (i.e., we can align the execution for an exponential configuration space while the number of nodes does not grow exponentially with the number of configurations) to most programs in the wild, because recent studies have shown that although there could be many options in a program, most options interact locally and thus interaction degrees are usually low~\cite{MWK+:ASE16, \samplingevidenceforlowdegree}.

\section{Applications Beyond Debugging}\label{sec:application}
\noindent
Understanding differences among executions can be useful beyond debugging.
In this section, we discuss three further potential applications of variational traces.

\paragraph{Detecting behavioral feature interactions} 
Detecting faults requires certain types of specifications (e.g., in form of test cases).
However, certain types of feature interaction faults are hard to detect if they do not trigger an exception but only have behavioral impacts.
For example, two features may interact in such a way that one feature suppresses the effect of another.
In previous work~\cite{SMN+:VaMoS18}, we presented \emph{feature interactions graphs} that help to identify which feature interact with each other and weather their interactions are suspicions (e.g., suppression).
These feature interactions graphs are based on an analysis of variational traces.

\paragraph{Understanding impact of load-time options}
It is hard to identify code that is affected by load-time options, especially due to implicit data flow.
Previous work used static taint analysis to detect code that depends on the selection of certain options~\cite{LKB:ASE14}.
However, it is hard to trace back why the code is affected as such information is lost in the analysis. 
With variational traces we can help to understand why certain parts of the code depends on the selection of options as we show causes of differences in the control flow. 

\paragraph{Understanding information flow}
Previous work compared executions to detect information flow, either using a similar technique to variational execution~\cite{AF:POPL12, AYF+:PLAS13, Schmitz:2016be, Yang:2015dl} or using multi execution~\cite{DP:SP10, KLZ+:SP12, KKS+:ASPLOS16, KKS+:ASPLOS15}.
Aligning executions allows to detecting potential leaks of secret data.
However, to understand why the information is leaked requires understanding why the executions differ.
Thus, again, variational traces can help understanding the causes of information flow by tracing data differences.

\section{Related Work}\noindent
We already discussed closely related work in the domains of automatic debugging~\cite{JHS:ICSE02, AZG:PRDC06,JHS:ICSE02,PO:ISSTA11,ZH:TSE02,Z:FSE02,SZ:ICSE13, SZ:FSE09,XSZ:PLDI08, GCK+:STTT06, AOH:ASE07} and feature interactions~\cite{CDS:ISSTA07, MKRGA:ICSE16,NL:CSUR11,KMS+13, SAG:ICSE17,RARF:JPF11,NKN:ICSE14, MWK+:ASE16,KBK:AOSD11,LKB:ASE14} in Section~\ref{sec:motivation}.
Our work combines ideas from both fields and builds specifically on the idea of sharing and coordinating multiple executions with variational execution~\cite{MWK+:ASE16, NKN:ICSE14, M14} and
thus sidesteps the challenges of trace alignment and full trace recording \cite{BGJ+:TC05, KC:ISPASS11} as discussed.\looseness=-1

\emph{Omniscient} or \emph{back-in-time debugging} allows exploring and debug a single execution~\cite{L:CoRR03, PTP:OOPSLA, BBR+:SIGHCI13, KM:TOSEM10, KM:ICSE08}.
In contrast to standard debuggers, back-in-time debuggers allow exploring information of previous parts of the execution.
To do so, they record full traces resulting in severe scalability challenges as they cannot predict which information is of interest.
In contrast, we provide a dynamic analysis that can decide on the fly which information is potentially relevant to explain the fault.
\looseness=-1

\emph{Symbolic execution} allows to explore all executions of a program for different inputs.
As discussed in Section~\ref{sec:varex}, symbolic execution usually does not share the executions after they are separated unless they incorporate ideas from variational execution~\cite{SNG+:FSE}.
The \emph{interactive verification debugger} is a tool to understand the symbolic execution of a program~\cite{HHB:ASE16:DEMO, HHB:ASE16}.
The tool visualizes the execution in a tree structure similar to Varviz.
However, as the symbolic execution never joins the tree structure gets large even for small programs.
In contrast, our variational trace provides a concise representation of many executions.\looseness=-1

\emph{Static and dynamic program slicing} are techniques to reduce a program to only the statements relevant for
understanding the state at a given program point~\cite{KL88, W:ICSE81}.
However, program slices are often large and cannot explain differences among multiple executions, especially execution omission bugs~\cite{ZTG+:PLDI}.
Differential slicing and dual slicing~\cite{JCC+:SP11, KKS+:ASPLOS15, SZ:ICSE13} compare the execution of two executions and reduces the comparison using program slicing.
In contrast to slicing approaches, we aim to explain the differences among many executions and only focus on the causes and differences in the state to keep the explanations concise.

\begin{sloppypar}
\emph{Multi execution} are approaches that synchronize (typically two) concrete executions.
These approaches enable analyses for information flow~\cite{DP:SP10, KLZ+:SP12, KKS+:ASPLOS16, KKS+:ASPLOS15}, configuration faults~\cite{SAF:SOSP07} and inconsistent updates~\cite{HC:ICSE13, MB:USENIX12,TXZ09}.
In contrast, our approach can compare a potentially exponential number of executions and helps to understand how the differences affect the program behavior.\looseness=-1
\end{sloppypar}

\section{Conclusion}\noindent
In this work, we propose variational traces to explain the runtime behaviors of inputs and interactions among them.
Using variational execution and specialized filters, we scale the generation of variational traces to the potentially exponential space of possible inputs.
To visualize variational traces, we provide an interactive Eclipse plugin called Varviz, which enables programmers to use variational traces for debugging interaction faults. 
In our user study, we show that users of Varviz outperform the users of the Eclipse debugger significantly in terms of understanding and time spent on debugging tasks.
Users of Varviz can focus on relevant parts of the programs quickly, without being distracted by irrelevant data and control flow decisions.
When compared with users who use standard Eclipse debugger, Varviz users can finish all the debugging and understanding tasks, using less than half of the time.
We further evaluate the size of variational traces on six highly configurable systems.
In general, the size of variational traces can get large, but our filters are effective in reducing the traces to a relatively small number of statements. 
Overall, our evaluation of effectiveness and scalability demonstrates that variational traces are useful in practice to understand differences among executions.\looseness=-1

\balance
\bibliographystyle{bibtex/myabbrv_noaddress_nomonth}
\bibliography{bibtex/MYfull,bibtex/literature}

\end{document}